\documentclass[10pt]{revtex4}
\usepackage{amsthm}
\usepackage{amssymb}
\usepackage{amsmath}
\usepackage{graphicx}

\def\dd{\mathcal D}
\def\zz{\mathcal Z}

\def\nab{\nabla}
\def\3nab{\tilde{\nabla}}

\def\p{\partial}

\def\nn{\nonumber}

\def\hsp5{\hspace{5mm}}
\def\case#1/#2{\textstyle\frac{#1}{#2}}

\def\be {\begin{equation}}
\def\ee {\end{equation}}
\def\ber {\begin{eqnarray}}
\def\eer {\end{eqnarray}}

\def\bc {\begin{center}}
\def\ec {\end{center}}
\newcommand{\hs}{\,-\,}

\begin{document}

\title{Gauge invariant perturbations of Scalar-Tensor Cosmologies I: The vacuum case}

\author{Sante Carloni$^{1}$,  Peter K. S. Dunsby$^{1,2}$, Claudio Rubano $^{3}$}

\address{$1.$ Department of Mathematics and Applied Mathematics, \\
University of Cape Town, 7701 Rondebosch, South Africa,}
\address{$2.$ South African Astronomical Observatory, Observatory 7925,
Cape Town, South Africa,}
\address{$3.$  Dipartimento di Scienze Fisiche e Sez. INFN di Napoli,
Universit\`{a} di Napoli "Frederico II", Compliesso Universitario
di Monte S. Angelo, Via Cynthia I-80126 Napoli (Italy).}

\begin{abstract}
The covariant gauge invariant perturbation theory of scalar
cosmological perturbations is developed for a general Scalar-Tensor
Friedmann-Lemaitre-Robertson-Walker cosmology in a vacuum. The
perturbation equations are then solved exactly in the long
wavelength limit for a specific coupling, potential and background.
Differences with the minimally coupled case are briefly discussed.
\end{abstract}

\maketitle

\section{Introduction} \label{sec:ch7intro}
In the last few years a tremendous advance in the accuracy of
cosmological observations has revolutionised our understanding of
the dynamics of the Universe. In particular, it has been realized
that the universe is evolving in a way that is incompatible with
what one would expect in a homogeneous and isotropic baryon
dominated universe. One way of dealing with this problem is to
suppose that the Universe is still homogeneous and isotropic but
dominated by a different form of energy density: the so called
Dark Energy (DE). Many different ideas regarding the nature of DE
have been proposed in the last few years, ranging from the
cosmological constant \cite{bi:cosmconst} to Quintessence
\cite{bi:quintessence}, Chaplygin gas \cite{bi:chaplygin} and
Phantom energy \cite{bi:phantom}, but so far none of them has
provided a fully satisfactory solution to the problem.

Another interesting approach is based on the hypothesis that on
cosmological scales gravity works in a slightly different way than
in General Relativity (GR). In this way DE acquires a geometrical
character, i.e. some of the observations can be explained by having
a different behaviour of gravity on cosmological scales, rather than
a new (and as yet undiscovered) form of energy density.

The idea that gravity might work differently on different scales is
suggested by many fundamental schemes such as quantum field theory
in curved spacetime \cite{bi:donoghue}, as well as the
compactification of internal spaces in multidimensional gravity
\cite{Les82,fuji maeda}, the low energy limit of string theory
\cite{Mtheory} and some braneworld models \cite{bi:chiba}.
Furthermore, GR has only been well tested on small scales and at low
energies, so currently we lack strong constraints in the other
regimes. In this respect cosmology is an ideal testing ground
because it offers an indirect way to test GR on scales and at
energies which are not achievable in Earth-bound laboratories, but
were realized throughout the history of the universe. Therefore,
investigations of cosmology in alternative theories of GR has a
double relevance; it contributes to the understanding of DE and
provides a testing ground for GR on different scales.

A key step in investigating the cosmology of an alternative theory
of gravity is the development of a full theory of cosmological
perturbations. This will make it possible to use some of the best
and most precise sets of observational data currently available,
such as the WMAP observations of the Cosmic Microwave Background
(CMB) anisotropies, allowing us to understand the viability of these
theories and constrain their parameters.

In this paper, we will develop the theory of linear scalar
perturbations for one of the most studied extensions of GR:
Scalar-Tensor theories of gravity. This will be done using the 1+3
covariant and gauge invariant approach, developed in
\cite{bi:ellis2} and successfully applied to GR in the presence of
minimally coupled scalars field \cite{bi:peterscalar}. Although this
formalism can be applied to perturbations of any spacetime, our work
will be focused only on {\em almost
Friedmann-Lemaitre-Robertson-Walker} (FLRW) universes which
correspond to standard linear perturbations of FLRW models (see
\cite{bi:EBH} for details) and in absence of standard matter. The
non-vacuum case will be treated in a forthcoming paper. Scalar field
dominated universes in the Scalar-Tensor framework have attained
prominence through the {\it Extended} and {\it Hyper-extended}
Inflation scenarios \cite{bi:inflext,bi:inflhyperext}, so an
analysis of the evolution of perturbations is important for
obtaining a complete description of structure-formation in these
models. In addition, the general formalism presented here could be
used in situations different from inflation in which a scalar field
dominates. As in \cite{bi:peterscalar}, emphasis is given here to
curvature perturbations  \cite{bi:hawking}, which are naturally
gauge invariant, rather than metric perturbations
\cite{bi:bardeen,bi:ks,bi:brand} which play no explicit role.
Bardeen's formalism has been applied to this situation in specific
cases of coupling, potential and background \cite{bi:salv1}. There
has also been quite an extensive analysis of perturbations in
non-standard theories of gravity given by Hwang in
\cite{bi:hwangall}. The aim of this paper is to give the general
perturbation equations for a general coupling, potential and (FLRW)
background using the full covariant approach to cosmologyand
choosing  a set of variables which are as much as possible of
straightforward physical interpretation.

The paper is organised as follows. In section \ref{sec:ch7prelim},
after a few preliminary remarks on Scalar-Tensor gravity, we set up
the formalism based on the natural slicing of the problem
$\{\phi=const\}$ and on its geometric characterisation through the
unit vector $u^a$, which is orthogonal to these surfaces. We also
characterise  the {\em effective} thermodynamics of the
non-minimally coupled field by treating it as an {\em effective}
fluid. In section \ref{sec:ch7dyn} we specify the general features
of the background and define the key gauge invariant dimensionless
variables. In section \ref{sec:ch7perteq} we present the general
perturbation equations, and, after specifying the coupling, the
potential and the background, we give an exact solution of the
perturbation equations in the large wavelength limit. In section 5,
we work out a specific example giving an exact solution in the long
wavelength limit i.e. on super-horizon scales.

Unless otherwise specified, natural units ($\hbar=c=k_{B}=8\pi G=1$)
will be used throughout this paper, Latin indices run from 0 to 3.
The symbol $\nabla$ represents the usual covariant derivative and
$\partial$ corresponds to partial differentiation. We use the
$-,+,+,+$ signature and the Riemann tensor is defined by
\begin{equation}
R^{a}{}_{bcd}=W_{bd,c}^{a}-W_{bc,d}^{a}+ W_{bd}^{e}W_{ce}^{a}-
W_{bc}^{f}W_{d f}^{a}\;,
\end{equation}
where the $W_{bd}^{a}$ is the Christoffel symbol (i.e. symmetric in
the lower indices), defined by
\begin{equation}
W_{bd}{}^{a}=\frac{1}{2}g^{ae}
\left(g_{be,d}+g_{ed,b}-g_{bd,e}\right)\;.
\end{equation}
The Ricci tensor is obtained by contracting the {\em first} and the
{\em third} indices
\begin{equation}\label{Ricci}
R_{ab}=g^{cd}R_{acbd}\;.
\end{equation}
Finally the Hilbert--Einstein action in presence of matter is
defined by
\begin{equation}
{\cal A}=\int d x^{4} \sqrt{-g}\left[\frac{1}{2}R+ L_{m}\right]\;.
\end{equation}

\section{Preliminaries} \label{sec:ch7prelim}
\subsection{Scalar Tensor Theories of Gravity}
\label{subsec:ch7scal}
Scalar-Tensor theories of gravity involve a new degree of freedom of
the gravitational interaction in a form of a scalar field
non-minimally coupled to the geometry. These theories were first
proposed as a modification of GR able to completely include Mach's
principle \cite{bi:BransDicke}. From a technical standpoint, this
can be achieved by allowing the Gravitational constant $G$ to vary
throughout the spacetime, and in particular to be a function of a
scalar field (so that frame effects are avoided). However, such a
scalar cannot be found within GR because they can contain first
derivatives of the metric and fall off more rapidly than $r^{-1}$.
As a consequence, $G$ is imagined to depend on an additional scalar
field $\phi$.

Like many of the extensions of GR, scalar-tensor theories appear in
many fundamental schemes and have been proposed as models for dark
energy because their cosmology naturally leads to the phenomenon of
{\em cosmic acceleration}, which is the characteristic footprint of
dark energy \cite{ScTnDarkEnergy,Claudio}.

The  most general  action for Scalar Tensor Theories of gravity is
given by (conventions as in Wald \cite{bi:wald})
\begin{equation}\label{eq:actionScTn}
\mathcal{A}=\int d x^{4}\sqrt{-g}\left[\frac{1}{2}F(\phi)R
-\frac{1}{2}\nab_a\phi\nab^a\phi -V(\phi)+\mathcal{L}_m \right]\;,
\end{equation}
where $V(\phi)$ is a general (effective) potential expressing the
self interaction of the scalar field and $\mathcal{L}_m$
represents the matter contribution.

Varying the action with respect to the metric gives the gravitational
field equations:
\begin{eqnarray}
&& \nn F(\phi)G_{ab}= F(\phi)\left(R_{ab}-\frac{1}{2}\,g_{ab}
R\right)=\\&&=T _{ab}^{m}+\nab_a\phi\nab_b\phi-g_{ab}
\left(\frac{1}{2}\nab_c\phi\nab^c\phi+V(\phi)\right)
+\left(\nab_b\nab_a - g_{ab}\nab_c\nab^c\right)F(\phi)\;,
\label{eq:einstScTn}
\end{eqnarray}
while the variation with respect to the field $\phi$ gives the
curved spacetime version of the Klein\hs Gordon equation
\begin{equation}
\nab_a\nab^a\phi+\frac{1}{2}F'(\phi) R -V'(\phi)=0\;, \label{eq:KG}
\end{equation}
where the prime indicates a derivative with respect to $\phi$.
Both these equations reduce to the standard equations for GR and a
minimally coupled scalar field when $F(\phi)=1$.

Equation (\ref{eq:einstScTn}) can be recast as
\begin{equation}
\label{eq:einstScTneff}
 G_{ab}=\frac{ T_{ab}^{m}}{F(\phi)}+T^{\phi}_{ab}=T^{(eff)}_{ab}\,,
 \end{equation}
 where $T^{\phi}_{ab}$ has the form
\begin{eqnarray}\label{eq:TenergymomentuEff}
T_{ab}^{\phi}=\frac{1}{F(\phi)}\left[\nab_a\phi\nab_b\phi-g_{ab}
\left(\frac{1}{2}\nab_c\phi\nab^c\phi+V(\phi)\right) +\nab_b\nab_a
F(\phi)- g_{ab}\nab_c\nab^cF(\phi)\right]\;. \label{eq:semt}
\end{eqnarray}
Provided  that $\phi_{,a} \neq 0$, equation (\ref{eq:KG}) also
follows from the conservation equations
\begin{equation}
\nab^bT_{ab}^{\phi}=0\;.
 \label{eq:cons}
\end{equation}
The fact that the gravitational field equation can be written in
this form is crucial for our purposes. In fact, the form of
(\ref{eq:einstScTneff}) allows us to treat scalar tensor gravity
as standard Einstein gravity in presence of two effective fluids
with energy momentum tensor $\frac{2 T_{ab}^{m}}{F(\phi)}$ and
$T^{\phi}_{ab}$. This implies that, once the effective
thermodynamics of these fluids has been studied, we can apply the
standard covariant and gauge-invariant approach to cosmology. In
this paper we will consider only the vacuum  $T_{ab}^{m}=0$ case,
leaving the non-vacuum case to a future paper.

\subsection{Kinematical quantities} \label{subsec:ch7kin}
In order to give a proper description of the kinematics of the
effective fluid associated with the scalar field we have to assign
a 4-velocity vector $u^a$ to the scalar field itself. Following
\cite{bi:peterscalar,bi:madsen} we assume that the {\it
momentum density} $\nab^a\phi$ is {\it timelike}:
\begin{equation}
\nab_a\phi\nab^a\phi<0\;, \label{eq:ass1}
\end{equation}
in the open region $A$ of spacetime we are dealing with. This has
two important consequences: Firstly ${\mathit \phi = const.}$
specifies well-defined surfaces in spacetime and secondly, these
surfaces are {\it spacelike} (as consequence of (\ref{eq:ass1})). We
can therefore choose the 4-velocity $u^a$ to be the normalised
timelike vector
\begin{equation}
 u^a\equiv - \frac{\nab^a\phi}{\psi}\;,~~~u^a u_a=-1\;,~~~\psi
\equiv \dot{\phi} = u^a\nab_a\phi = (-\nab_a\phi
\nab^a\phi)^{1/2}\;, \label{eq:u}
\end{equation}
where $\psi=\dot{\phi}$ denotes the magnitude of the
momentum density (simply momentum from now on). The
vector (\ref{eq:u}) is timelike because it is parallel to the
normals of (i.e. orthogonal to) the surfaces $\{\phi=const.\}$ and
it is also unique because, since these surfaces are well defined,
their normals are unique.

Note that this definition might not appear well posed in general
because $\phi$ could oscillate and change sign even in an
expanding universe. However, equation (\ref{eq:ass1}) and the
regularity of the behaviour of $\phi$ (see \cite{bi:peterscalar}
for more details) ensures the possibility of an extension by
continuity.

Once the velocity field has been chosen, the projection tensor
into the tangent 3-spaces orthogonal to the flow vector can be
defined as:
\begin{equation}
h_{ab}  \equiv g_{ab}+u_au_b\; \Rightarrow h^a{}_b h^b{}_c=h^a{}_c\;,
~h_{ab}u^b=0\;,
\end{equation}
and the derivation of the kinematical quantities can be obtained by
splitting the covariant derivative of $u_a$ into its irreducible
parts \cite{bi:ellis2}:
\begin{equation}
\nab_b u_a=\3nab_b u_a-A_a u_b\;, ~~~\3nab_b u_a=\frac{1}{3}\Theta
h_{ab} +\sigma_{ab}+\omega_{ab}\;, \label{eq:dec}
\end{equation}
where $\3nab_a$ is the totally projected covariant derivative
operator orthogonal to $u^a$, $A_a = \dot{u}_a$ is the acceleration
($A_bu^b=0$), $\Theta$ is the expansion parameter, $\sigma_{ab}$ the
shear ($\sigma^a{}_a = \sigma_{ab}u^b=0$) and $\omega_{ab}$ is the
vorticity ($\omega_{ab} =\omega_{[ab]}$, $\omega_{ab}u^b=0)$.

Important consequences of the choice (\ref{eq:u}) are
\begin{equation}
\3nab_a\phi=0, \label{eq:gfi}
\end{equation}
i.e. the surfaces $\{\phi=const.\}$  are the 3-spaces orthogonal
to $u^a$, and
\begin{eqnarray}
\Theta& = & -\3nab_a(\frac{\nab^a\phi}{\psi})
=-\frac{1}{\psi}\left[F'(\phi) R-V'(\phi)+\dot{\psi}\right] \neq0\;, \label{eq:exp}\\
\sigma_{ab} & = & - \frac{1}{\psi}\,h_a{}^c h_b{}^d \nab_{(c}
\left[
\nab_{d)}\phi \right] + \frac{1}{3} h_{ab} \nab_c (\frac{\nab^c\phi}{\psi}) \neq0\;,\\
A_a & = & -\frac{1}{\psi}\,\3nab_a\psi=-\frac{1}{\psi}(\nab_a\psi
+u_a\dot{\psi})\neq0\;, \label{eq:acc2}
\end{eqnarray}
where the last equality in (\ref{eq:exp}) follows on using the
Klein\hs Gordon equation (\ref{eq:KG}). We can see from
(\ref{eq:acc2}) that $\psi$ is an {\em acceleration potential} for
the fluid flow \cite{bi:ellis2}. Note also that, as consequence of
(\ref{eq:gfi}), the vorticity vanishes:
\begin{equation}
\omega_{ab}=-h_a{}^c
h_b{}^d\nab_{[d}\left(\frac{1}{\psi}\nab_{c]}\phi\right)=0\;,
\end{equation}
which implies that $\3nab_a$ is the actual covariant derivative
operator in the 3-spaces ${\mathit \phi=const}$ . Finally, it is
useful to introduce a scale length $ a $ along each flow\hs line so
that
\begin{equation}
\frac{\dot{ a }}{ a }\equiv \frac{1}{3} \Theta = H\;,
\end{equation}
where $H$ is the usual Hubble parameter if the Universe is
homogeneous and isotropic.
\subsection{The non-minimally coupled scalar field as an effective imperfect fluid} \label{subsec:ch7imp}
Using the definition of the 4-velocity given above, we can deduce the
thermodynamical properties of the effective fluid described by the
non minimally coupled  scalar field. The general form of the
energy momentum tensor (\ref{eq:semt}) is:
\begin{equation}
T_{ab}^{\phi}=\mu_{\phi} u_a u_b +p_{\phi}
h_{ab}+q_a^{\phi}u_b+q_b^{\phi}u_a+\pi_{ab}^{\phi}\;,
\label{eq:pf}
\end{equation}
where the energy density $\mu_{\phi}$ and pressure $p_{\phi}$ of
the scalar field ``fluid'' are:
\begin{equation}\label{eq:ed}
\mu_{\phi} =\frac{1}{F(\phi)}\left[\frac{1}{2}\psi^2 +V(\phi)
-\Theta\dot{F}(\phi)\right]\;,
\end{equation}
\begin{equation}\label{eq:p}
p_{\phi}=\frac{1}{F(\phi)}\left[\frac{1}{2}\psi^2 -V(\phi)
+\left(\ddot{F}(\phi)+\frac{2}{3}\Theta\dot{F}(\phi)\right)\right]\;,
\end{equation}
and the energy flux and anisotropic pressure are given by:
\begin{equation}
q_a^{\phi}=\frac{\dot{F}(\phi)}{F(\phi)}\; A_a \label{eq:qflux}\;,
\end{equation}
\begin{equation}
\pi_{ab}^{\phi}=-\frac{\dot{F}(\phi)}{F(\phi)}\;\sigma_{ab}\;.
\label{eq:anisp}
\end{equation}
Since the anisotropic pressure $\pi_{ab}$ \footnote{Only
scalar fields will be considered in the following sections,
so we drop the subscript from these quantities. } and the heat flux $q_a$
are different from zero, the effective fluid associated with a non-minimally
coupled scalar field is, in general, {\em imperfect}.
The form of (\ref{eq:qflux}) and (\ref{eq:p}) shows explicitly
that this is a direct consequence of the non-minimal coupling and
that, when $\phi$ is minimally coupled (i.e. $F(\phi)=1=constant$)
we have $q_a=0=\pi_{ab}$. This is in agreement with the results of
\cite{bi:peterscalar}, in which a minimally coupled scalar field is
considered to be a perfect fluid. It is also worth noting that, since
$\pi_{ab}$ and $q_a$ depend on the acceleration and the shear, when we
consider FLRW universes, where these quantities vanish, the effective
fluid behaves like a perfect fluid \footnote{Strictly speaking this is true
only if the observers are comoving with the fluid in the background. An observer
moving relative to a perfect fluid will measure a heat flux
and an anisotropic pressure due to a frame effect
\cite{bi:KingEllis}.}. However  this is not true in the perturbed
universe.

The twice contracted Bianchi Identities lead to
the energy and momentum conservation equations:
\begin{equation}
\dot{\mu} +\Theta(\mu
+p)+\pi_{ab}\sigma^{ab}+\3nab_{a}q^a+a_aq^a=0\;, \label{eq:encon}
\end{equation}
and
\begin{equation}
A_a(\mu+p)+\3nab_ap+h^c{}_{a}\left(\3nab_b
\pi^b{}_{c}+\dot{q}_c\right)+\left(\sigma^b{}_a+ \frac{4}{3}\Theta
h^b{}_a\right)q_b=0\;. \label{eq:momcon}
\end{equation}
If we now substitute $\mu$ and $p$ from (\ref{eq:ed}) and
(\ref{eq:p}) into (\ref{eq:encon}), we obtain, as expected, the
Klein\hs Gordon equation (\ref{eq:KG}):

\begin{equation}
\ddot{\phi} +\Theta \dot{\phi}-\frac{1}{2}F'(\phi) R +V'(\phi)=0\;,
\label{eq:KG2}
\end{equation}
an exact ordinary differential equation for $\phi$ in any space\hs time
with the choice (\ref{eq:u}) for the four\hs velocity. With the same
substitution, (\ref{eq:momcon}) becomes an identity for the
acceleration potential $\psi$.

In our calculation the ratio between the pressure $p$ and energy
density $\mu$ of the effective fluid defined above is given by
\begin{equation}
 w=-1+ \frac{1}{\mu F(\phi)}\left[
\psi^2+\frac{1}{3}\Theta\dot{F}(\phi)+\ddot{F}(\phi)\right]\;.
\label{eq:wphi}
\end{equation}
This is a useful parameter of the theory, but should not be confused with the
barotropic factor typical of perfect fluids. In the same way,
the quantity $\frac{\dot{p}}{\dot{\mu}}$ will be denoted by
\begin{equation}
c_s^2=\frac{\dot{w}}{w} \frac{\mu}{\dot{\mu}}+w\;, \label{eq:sos}
\end{equation}
even if it does not represent the proper speed of sound.
\section{Gauge\hs invariant perturbations and
their dynamics} \label{sec:ch7dyn}
\subsection{Background dynamics}\label{subsec:ch7grav}
The first step in developing  a theory for the evolution of
cosmological perturbations is the definition of the background. For
this purpose it is useful to employ the {\em 1+3 covariant approach}
to cosmology \cite{bi:ellis2}. This method consists of writing the
Einstein Field Equations and their integrability conditions as a
system of six exact evolution and six constraint equations ({\em 1+3
equations}) for a set of covariantly defined quantities (which
include the expansion, the shear and the vorticity, already defined
in section \ref{subsec:ch7scal}). The advantage of such a
re-parametrisation is that the treatment of the exact theory is
considerably simplified, making it easier to find background
solutions, even in more complicated cosmological models (like the
Bianchi spacetimes). In addition, many of these 1+3 variables can be
shown to be gauge-invariant \cite{bi:BDE} and it is these quantities
that form the building blocks of our gauge-invariant perturbation
theory.

In this paper we limit our analysis to the study of perturbations of
spatially homogeneous and isotropic cosmological models
(FLRW models), in which only a spatially homogeneous non-minimally
coupled classical scalar field is present. This means that in the
background we have
 \begin{equation}
 \sigma =\omega= 0\;, ~~\3nab_a f=0\;,
 \label{eq:rwcond1}
 \end{equation}
 where $f$ is any scalar quantity; in particular
 \begin{equation}
 \3nab_a\mu =\3nab_a p=0 ~~\Rightarrow~~\3nab_a \psi = 0\,, ~A_a = 0  \;.
 \label{eq:rwcond3}
 \end{equation}
In this way the 1+3 equations reduce to a system of only two
equations: the Raychaudhuri equation
  \begin{equation}
\dot{\Theta}+\frac{1}{3}
\Theta^2+\frac{1}{2}\left(\mu_{\phi}+3p_{\phi}\right)= 0\;,
\label{eq:ray}
\end{equation}
and the Gauss\hs Codazzi equation
\begin{equation}
\tilde{R} = 2\left[- \frac{1}{3} \Theta^2+\mu_{\phi}\right]\;,
\label{eq:GK}
\end{equation}
where $\tilde{R}$ represents the 3-ricci scalar of the ${\mathit
\phi=const}$ surfaces, and is closed by the conservation equation
(\ref{eq:encon}).

Expressing  (\ref{eq:encon}), (\ref{eq:KG2}), (\ref{eq:GK}),
(\ref{eq:ray}) in terms of the Hubble parameter $H$ and the
momentum variable $\psi$, we obtain  the background (zero\hs order)
equations:
 \begin{equation}
 3\dot{H}+3H^2 =-\frac{1}{F(\phi)}\left[\psi^2
-V(\phi)+\frac{1}{2}\Theta\dot{F}(\phi)+\frac{3}{2}\ddot{F}(\phi)\right]\,
,\label{eq:e1b}
 \end{equation}
 \begin{equation}
 3H^2+3K=\frac{1}{F(\phi)}\left[\frac{1}{2}\psi^2
+V(\phi)-\Theta\dot{F}(\phi)\right] \, ,
 \label{eq:e1a}
 \end{equation}
 \begin{equation}
 \dot{\psi}+3H\psi-\frac{1}{2}F(\phi) R +V'(\phi)= 0\;\Leftrightarrow
 \label{eq:e1c}
 \end{equation}
$$
  \dot{\mu} + 3H\left(\psi^2+ \frac{1}{3}\Theta\dot{F}(\phi)+\ddot{F}(\phi)\right) =0 \; ,
$$
 where all variables are functions of cosmic time $t$ only, and
 \begin{equation}
 \tilde{ R} = 6K/ a ^2, ~K = const
 \end{equation}
as required by the properties of homogeneous and isotropic
spacetimes.
\subsection{Gauge\hs invariant perturbation variables}\label{subsec:ch7var}
\subsubsection{Spatial gradients}
Using the 1+3 variables we can define exact quantities that
characterise inhomogeneities in any space\hs time, and also derive
exact non\hs linear equations for them. For example, to
characterise the energy density perturbation, the natural variable
is $X_a=\3nab_a\mu$. This vector represents any spatial
variation of the energy density $\mu$ (i.e. any over-density or
void) and it is in principle directly measurable \cite{bi:EB}.
However, a more suitable quantity to describe density perturbation
is
\begin{equation}\label{eq:vardefD}
\dd_a=\frac{ a }{\mu}X_a =\frac{ a }{\mu}\3nab_a\mu\;,
\end{equation}
where the ratio $X_a/\mu$ allows one to evaluate the magnitude of
density perturbations relative to the background energy density and
the presence of the scale factor $a$ guarantees that it is
dimensionless and comoving in character. The magnitude of $\dd_a$ is
closely related with the more familiar quantity
$\displaystyle{\frac{\delta\mu}{\mu}}$, the difference being that
$\dd=(\dd_a\dd^a)^{1/2}$ represents a real spatial fluctuation
instead of a fictitious gauge dependent one. In fact, it is easy to
show that the Bardeen variable $\epsilon_m$ which corresponds to
$\displaystyle{\frac{\delta\mu}{\mu}}$ in the comoving  gauge, is
the scalar harmonic component of $\dd_a$ (i.e. its scalar
``potential").

Other important quantities relevant to the evolution of density
perturbations are
\begin{equation}
 \zz_a \equiv  a
\3nab_a\Theta\;,
 ~~ C_a \equiv   a^3 \3nab_a \tilde{R}\;, \label{eq:vardefZC}
\end{equation}
which represent, the spatial gradient of the expansion and the
spatial gradient of the 3-Ricci scalar respectively. These
vectors, together with $\dd_a$, vanish in the background FLRW model
and are therefore gauge-invariant by the Stewart-Walker
lemma \cite{bi:stewart}.

At this point one could  derive the exact non\hs linear evolution
equations for density perturbations from the definitions
above and the equations (\ref{eq:encon}), (\ref{eq:KG2}),
(\ref{eq:GK}), (\ref{eq:ray}). However, since in our case we
are considering an effective fluid representing the non-minimally
coupled scalar field, we may want to characterise directly the
inhomogeneity of $\phi$. This cannot be done using $\3nab_a\phi$
because our frame choice (\ref{eq:u}) implies that the spatial
gradient $\3nab_a\phi$ identically vanishes in any space\hs time.
As a consequence (and just like the minimally coupled case
\cite{bi:peterscalar,bi:barplus}), the natural gauge invariant
perturbation variable for the inhomogeneities of the scalar field
in our approach is the spatial variation of the momentum
$\3nab_a\psi$. Following the same reasoning used for $\dd_a$ we
can define the dimensionless gradient
\begin{equation}
\Psi_a\equiv \frac{ a }{\psi}\3nab_a\psi\;. \label{eq:psidef}
\end{equation}
Since an exact FLRW Universe is characterised by the conditions
(\ref{eq:rwcond1}), (\ref{eq:rwcond3}), $\Psi_a$
vanishes identically in the background and therefore is {\it
gauge\hs invariant}. In addition, by comparing (\ref{eq:acc2}) and
(\ref{eq:psidef}), we see that $\Psi_a$ is proportional to the
acceleration ($a_b=\Psi_{b}/ a ^2$). Thus, this quantity can be
interpreted as a gauge invariant measure of the spatial variation
of proper time along the flow lines of $u^a$ between two surfaces
$\phi=const.$ \cite{bi:barplus,bi:ellis1}.

\subsubsection{Linearization}

Since the variables ($\dd_a$, $\zz_a$, $C_a$, $\Psi_a$) are
completely general, solving their evolution equations in full would
be equivalent to solving the full Einstein equations, a very
difficult task. On the other hand, the current observations suggests
a universe which is very close to FLRW. This means that we can limit
our analysis to situations where the real universe does not differ
too much from the background, i.e. when the magnitude of the
quantities in (\ref{eq:rwcond1}), (\ref{eq:rwcond3}),
(\ref{eq:vardefD}) and (\ref{eq:vardefZC}) are small. This can be
achieved by treating the quantities that do not vanish in the
background as {\em zero order} and those that vanish (and are hence
gauge-invariant) as {\em first order},  retaining only terms which
are first order in the gauge-invariant perturbation variables.

In our specific case (background given by FLRW universe + a scalar
field), the zero order quantities are given by $\Theta$, $\mu$, $p$
and the first order ones are $\sigma_{ab}$, $q_a$,$\pi_{ab}$,
$\dd_a$, $\zz_a$, $C_a$, $\Psi_a$. So for example, when we linearise
the equations, we drop the vector $\sigma_{ab}\dd^{a}$ because it is
second order, but we keep the quantity $\Theta\dd_a$, where it is
understood that $\Theta$ is the expansion in the background.

In the linear approximation we find two relations between the
perturbation variables defined above:
\begin{equation}\nn
C_a=-\frac{4}{3}\Theta  a ^2 \zz_a +2 \mu  a ^2 \dd_a\;,
\label{eq:constr}
\end{equation}
obtained by taking the gradient of (\ref{eq:GK}) and
\begin{equation}\nn
a^2\mu\Theta\left[2F(\phi)-3\dot{F}(\phi)\right]\dd_a=
4a^2\left(\psi^2+\Theta
\dot{F}(\phi)\right)\Psi_a-3\dot{F}(\phi)C_a\;,
\end{equation}
obtained from the definition of the effective energy density. Using
these equations we can express the heat flux $q_a$ and the
anisotropic pressure $\pi_{ab}$  in terms of these variables. For
the heat flux we have
\begin{equation}\nn\label{eq:heatfluxUpsi}
    q_b=\frac{\dot{F}(\phi)}{F(\phi)}\; a_b =\frac{\dot{F}(\phi)}{F(\phi)}\;\frac{\Psi_a}{a ^2}\;,
\end{equation}
where we have also used the relation between acceleration and
$\Psi_a$. For the gradient of the anisotropic pressure we have
\begin{eqnarray}\label{eq:anipressUpsi}
   \nn \3nab^{b}\pi_{ab}=\frac{\dot{F}(\phi)}{F(\phi)}\,\left(\frac{C_a}{2\,{a}^4\,\Theta
}- \frac{\mu\,D_a }{{a}^2\,\Theta } -
\frac{\dot{F}(\phi)\,\Psi_a}{F(\phi)\,{a}^2} \right)\;,
\end{eqnarray}
where we have also used the linearised vorticity free shear
constraint
\begin{equation}\label{shear constraint}
    \frac{2}{3}Z_a-a\3nab^b\sigma_{a b}-a q_a=0\;.
\end{equation}

\subsubsection{Scalar gauge\hs invariant variables}
The vectors ($\dd_a$, $\zz_a$, $C_a$, $\Psi_a$) contain information
about the evolution of the energy density perturbations that is not
necessarily related with the formation of local inhomogeneities
\cite{bi:EBH}. The relevant parts can be extracted using a local
decomposition. For example in the case of the gradient of $\dd_a$ we
have:
\begin{equation}
 a \3nab_b\dd_a \equiv\Delta_{ab}= \frac{1}{3} h_{ab}\Delta +\Sigma_{ab}
+W_{ab}\;,\label{eq:dec1}
\end{equation}
where
\begin{equation}
 \Sigma_{ab}\equiv \Delta_{(ab)}-
\frac{1}{3} \Delta h_{ab}\;,~~W_{ab}  \equiv \Delta_{[ab]}\;,
\end{equation}
are the symmetric, trace\hs free and anti\hs symmetric
parts of $\Delta_{ab}$ respectively. The first represents spatial
variations of $\dd_a$  associated with change in the spatial
anisotropy pattern of this gradient field and it can be related
to the formation of non-spherically symmetric structures. The
second one represents spatial variations of $\dd_a$ due to rotation
of the density gradient field. Using the definition
of $W_{ab}$ it is easy to see that this tensor is  proportional to
the vorticity \cite{bi:EBH} and therefore vanishes for our choice
of 4-velocity $u^a$. The first term $\Delta$ in (\ref{eq:dec1}) is the
trace of $\Delta_{ab}$. It represents the spherically symmetric spatial
variation of the energy density $\mu$ and is the variable most closely
associated with matter clumping. The evolution of this variable
will be the focus of this paper.

In the same way one can obtain  the corresponding decomposition of
the variables (\ref{eq:vardefZC}) and consider only the trace of
their gradient. In this way  we have the set of scalar gauge
invariant-variables:
\begin{equation}
\Delta \equiv   a \3nab^a\dd_a\;, ~~ \zz \equiv   a
\3nab^a\zz_a\;, ~~ C  \equiv   a \3nab^a C_a\;,~~ \Psi\equiv  a
\3nab^a\Psi_a\;, \label{eq:cdef}
\end{equation}
respectively giving the energy density, expansion, 3-curvature
scalar perturbations and the spatial distribution of the momentum
$\Psi$. Constraint equations analogous to (\ref{eq:constr}) and
(\ref{eq:DeltaUpsiC}) then relates $\zz$, $C$, $\Delta$  and
$\Psi$:
\begin{equation}
C=-\frac{4}{3}\Theta  a ^2 \zz +2\mu  a ^2\Delta\;,
\label{eq:constr1}
\end{equation}
\begin{equation}\label{eq:DeltaUpsiC}
a^2\mu\left[2\Theta F(\phi)-3\dot{F}(\phi)\right]\Delta= 2a^2
\Theta\left(\psi^2+\Theta
\dot{F}(\phi)\right)\Psi-3\dot{F}(\phi)C\;.
\end{equation}
In the minimally coupled case ($F(\phi)=$ constant) this last
equation reduces to
\begin{equation}
\Delta=(w+1) \Psi\;, \label{eq:delta1}
\end{equation}
which is equivalent to the result found in \cite{bi:peterscalar},
and we conclude that, in this case, $\Delta$ characterises
scalar field clumping. However, this is not true in
general for a non-minimally coupled
scalar field because (\ref{eq:DeltaUpsiC}) is related to
the variable $C$, the details of the coupling function $F$ and its derivative,
as well as specific features of the background like $\Theta$ and
the sign of $\phi$.

This is not surprising: $\Delta$ represents the spherically
symmetric clumping of the effective fluid described by the stress
energy tensor (\ref{eq:TenergymomentuEff}) and, in general, it has
no direct relation with the inhomogeneities of the scalar field.
However, the behaviour of this quantity can still be interesting.
For example, in the context of the inflationary scenarios, even if
the perturbations of the scalar field are described by $\Psi$,
only $\Delta$ is connected with the generation of the seeds of
structure formation.

Finally, for later convenience, it is useful to
define the vector
\begin{equation}\label{eq:BardeenPHI}
    \Phi_a=\mu a^{2} \dd_a
\end{equation}
and its comoving divergence $\Phi$ defined in \cite{bi:peterscalar}.
This quantity is equal, up to a constant, to the combination
\begin{equation}
    \Phi_N=\frac{1}{2}\left(\Phi_A-\Phi_H\right)
\end{equation}
of the Bardeen potentials $\Phi_A$ and $\Phi_H$ and represents the
Newtonian gravitational potential. We will consider the
evolution of $\Phi_N$ in the specific example given in the last section.

\subsection{Entropy perturbations} \label{subsec:ch7entr}
As we have seen, even if the effective fluid in our background
takes on the form of a perfect fluid, this is not the case in the
perturbed universe. Furthermore, its equation of state does not
have the simple barotropic form $p=p(\mu)$, so that the
perturbations are in general {\it non-adiabatic} and we
need to introduce another variable to characterise the spatial
variation of the entropy density $s$. The form of this variable
is obtained by taking the spatial gradient of the equation of
state $p=p(\mu,s)$:
\begin{equation}\label{eq:grad-state}
 a \3nab_a p=\left.\frac{\p p}{\p\mu}\right|_{s} \!\!  a\, \3nab_a\mu
+\left.\frac{\p p}{\p s}\right|_{\mu} \!\!  a\, \3nab_a s\;,
\end{equation}
where we have the usual thermodynamic partial derivatives at
constant density and entropy \footnote{Note that, if we have a
perfect fluid in the background and we neglect the effect of bulk
viscosity, entropy is constant along flow lines and  the ratio
$\dot{p}/\dot{\mu}$ corresponds to the speed of sound so that the
definition of the speed of sound (\ref{eq:sos})\, coincides with the
standard thermodynamic definition.}. The second term in the equation
above represents the comoving gradient of the entropy $s$:
\begin{equation}
{\cal E}_a =\frac{ a }{p}\left.\frac{\p p}{\p s}\right|_{\mu}\,
\3nab_a s\;, \label{eq:ep}
\end{equation}
and corresponds to an effective entropy perturbation variable. Defining the
comoving fractional pressure gradient $\displaystyle{P_a\equiv
\frac{ a }{p}\3nab_a p}$, equation (\ref{eq:grad-state}) can be
written as
\begin{equation}\label{eq:defEntropy_a}
   {\cal E}_a=\frac{c_s^2 }{w}\dd_a - P_a \;,
\end{equation}
which means that
\begin{equation}
{\cal E}_a=a \frac{\3nab_a w}{w}-\frac{c_s^2-w}{w}\dd_a\;,
\end{equation}
where we have used the definition of the effective barotropic
factor $w=p/\mu$. Note that, ${\cal E}_a$ is proportional to
$\dd_a$ if $w=const$ and it is zero in the case of a
perfect fluid ($c_s^2=w$). The definition of ${\cal E}_a$ is very
close to the definition of the Bardeen variable $\eta$
\cite{bi:bardeen}, i.e., representing the difference between
pressure and density perturbations when the perturbations
are not adiabatic.

As for $\nabla_a\dd_b$, we can decompose $\nabla_a {\cal E}_b$ in
such a way to isolate only the scalar part that is relevant for our
purposes:
\begin{equation}\label{eq:defEntropy}
{\cal E}=a^2 \frac{\3nab^{2} w}{w}-\frac{c_s^2-w}{w}\Delta\;.
\end{equation}
Substituting for $\3nab^{2} w$ from (\ref{eq:wphi}), we obtain,
after a lengthy calculation,
\begin{eqnarray}\label{eq:EpsilonExpanded}
 {\cal E}& =&\nn\frac{\dot{F}}{F\,w\,\mu }\,\dot{\Psi}+
\frac{\dot{F}}
   {4\,\mu\,w\,F\,{a}^2\,\Theta}\;C +
  \frac{ 6\,{\psi }^2-\Theta \,\dot{F}+ 3\,\ddot{F} - 3\,{\psi }^2\,F''
       }{3\,w\,\mu \,F}\;\Psi
\\  &+& \left[\frac{
      F\, \left({\Theta}^2\, \dot{F} - 3\, \mu\, \dot{F} -
          3\, \Theta\, \ddot{F} +
           18\psi\, \dot{\psi}\, +9\dddot{F} \right)-3\, \dot{F}\,
    \left(3\, {\psi}^2 - \Theta\, \dot{F} + 3\, \ddot{F}\right)}{9\,
    w\, \Theta\, \mu\, {F}^2\, \left(1 + w \right)}\right]\, \Delta\;.
\end{eqnarray}
This will be will be used later in the derivation of the
perturbation equations.
\section{Perturbation equations} \label{sec:ch7perteq}
\subsection{First\hs order linear equations}
We are now ready to write down the linear evolution equations for
the scalar gauge-invariant variables defined above. Since we have
four perturbation variables $\Delta$, $\Psi$, $C$ \footnote{In this case as in the
minimally coupled one the variable $$\tilde{C} \equiv  a \3nab^a
\tilde{C}_a = C -\frac{4K}{w+1}\Delta\;,$$ is not a conserved
quantity for scales larger than Hubble radius
\cite{bi:peterscalar}. For this reason we will not include it in
our discussion.} and $\zz$, we will have a system of four
differential equations. Using the constraints (\ref{eq:constr1})
and (\ref{eq:DeltaUpsiC}), it can be reduced to a system of only two
equations. However, in the following we will consider equations
for the variables $\Psi$, $\Delta$, $C$, i.e., we do not
use the constraint (\ref{eq:DeltaUpsiC}). This choice is motivated
for the sake of simplicity: the system for these three variables
is  much easier to  manage than the general equations for two
variables only. The choice of $C$ as our third variable is, of course,
purely arbitrary: one could, for example, couple the first order
evolution equation for $\Delta$ and $\Psi$ to the equation for $\zz$,
but we choose $C$ because it makes it easier to make a direct
comparison with the minimally coupled scalar field and barotropic
fluid cases.

Starting from the definition of the gauge invariant variables
given above and using the fact that at linear order
$a\3nab_{a}\dot{X}_b=a\left(\3nab_{a}X_b\right)^{\centerdot}$\;
holds, we obtain the following system of equations:
\begin{eqnarray}
 \dot{\Delta}&=&\frac{3}{4}\frac{(w+1)  C}{ a ^2\Theta}
+\left[w
             \,\Theta - \frac{3}{2}\mu(w+1)
             \Theta^{-1}\right]\Delta-\frac{a}{\mu}\3nab^2\3nab_a
             q^a+\frac{a}{\mu}\Theta {\cal S}/:,
\label{eq:deltaeq1}
\end{eqnarray}
\begin{eqnarray}
 \dot{C} &=& \nn\frac{4\,{a}^2\,\Theta \, {{c_s}}^2}{3\,\left(
1 +
 w \right) }\left(\,\3nab^{2}+\frac{3 K}{{a}^2}\right)\Delta
+ \frac{4\,\,{a}^2\,w\,\Theta}
            {3\,\left( 1 + w \right)}\left(\3nab^{2}+\frac{3 K}{a^2}\right){\cal E}\nonumber\\
 &+& \frac{4\,{a}^4\,\Theta }{3\,\left( 1 + w \right)\,\mu
}\left(\,\3nab^{2} +\frac{3 K}{{a}^2}\right){\cal S} -2\,{a}^4
\,\3nab^{2}\3nab_a q^a  + 3K\left(\frac{\,C}{a^2\,\Theta }
            - \frac{2\,\,\Delta\,\mu}{\Theta}\right)\;, \
\label{eq:cdot1}
\end{eqnarray}
where
\begin{equation}
{\cal S}=\3nab_a \dot{q}^a+\frac{4}{3}\Theta\3nab_a
q^a+\3nab_a\3nab_b\pi^{a b}\;,
\end{equation}
together with an equation for $\Psi$ given by
\begin{eqnarray}
 \dot{\Psi}&=&\nn\frac{6\,F\,\Theta \,\mu\,\left( 1 + w \right)
      - 9 \,\mu\,\dot{F}\,\left( 1 + w \right)
     + 6\,\Theta \,\ddot{F}}{8\,{a}^2\,{\Theta }^2 \,\left( \psi^{2}  - \Theta \,\dot{F} \right) }\,C \nn \\
 &-&\frac{{\psi  }^2\left( 2{\Theta  }^2 -
     3\,\left( 1 + 3\,w  \right) \,\mu \right) +
      12\,\Theta  \,\psi  \,\psi '  +
      6\dot{F}\Theta \,\mu \left( 1 + 3\,w  \right)   -
      6{\Theta  }^2\,\ddot{F}  }{6\,\Theta  \,\left( \psi^{2}  - \Theta \,\dot{F} \right) }\,\Psi \nn \\
 &+&\left(
       \frac{ F  \,
         \left( 2\,{\Theta  }^2 +  3\,\left( 2 + 3\,w \right)\,\mu   \right) }
       {3\,\Theta  \,\left( \psi^{2}  - \Theta \,\dot{F} \right) }-\frac{\dot{F} \left(9\,\left( 1 + w  \right) \,\mu-2\,{\Theta  }^2\right) -
       6\,\Theta   \ddot{F} }{4\,{\Theta  }^2 \,\left( \psi^{2}  - \Theta \,\dot{F} \right) }
       \right)\,\Delta\,\mu \nn \\
 &+&\frac{{a }^2 \,
       \left( 2\,F\,\Theta   + 3\,\dot{F} \right) }{2\,\left( \psi^{2}  - \Theta \,\dot{F} \right)}{\cal S}
-\frac{{a }^2\,\dot{F}}{\left( 1 + w  \right) \,\mu  \,\left(
        \psi^{2} - \Theta \,\dot{F} \right) }\,\left( \3nab^2 +
      \frac{3\,k }{{a }^2} \right)\,{\cal S} \nonumber \\
 &-& \frac{F\,{a }^2} {\psi^{2}  - \Theta \,\dot{F}
       } \,\3nab^2 \3nab_a \dot{q}^a-\frac{w\,\dot{F}}{\left( 1 + w
      \right)\,\left( \psi^{2}  - \Theta \,\dot{F} \right)}\,\left( \3nab^2 +
      \frac{3\,k }{{a }^2} \right){\cal E} \nonumber \\
 &-& \frac{ {{c_s}}^2   \,\dot{F}}{\left( 1 + w  \right)
       \,\left( \psi^{2}  - \Theta \,\dot{F} \right) }\,\left( \3nab^{2}
      - \frac{3\,k}{{a }^2} \right)\,\Delta +\frac{9\,k \,\dot{F}}
      {4\,{a}^2\,{\Theta }^2\,\left( \psi^{2}  - \Theta \,\dot{F} \right)
      }\,\left( \frac{C}{{a}^2 } -2\,\mu\,\Delta
\right)\,,\label{eq:upsi1}
\end{eqnarray}
and the constraint (\ref{eq:DeltaUpsiC}):
\begin{equation}
\nn a^2\mu\left[2\Theta F-3\dot{F}\right]\Delta=
4a^2\Theta\left(\psi^2-\Theta \dot{F}\right)\Psi-3\dot{F}C\;.
\end{equation}
These equations reduce to those for a minimally coupled
scalar field when $F(\phi)=1$. Substituting for the heat flux, the
anisotropic pressure and the entropy in (\ref{eq:deltaeq1}),
(\ref{eq:cdot1}) and (\ref{eq:upsi1}) we obtain:
\begin{eqnarray}\label{eq:SysPertFin}
\nn\dot{\Delta}=&&\mathcal{A}_{\Delta}\,C+
\mathcal{B}_{\Delta}\,\Delta+ \mathcal{C}_{\Delta}\,\Psi+
\mathcal{D}_{\Delta}\, \3nab^{2}C
+\mathcal{E}_{\Delta}\,\3nab^{2}\Delta+ \mathcal{F}_{\Delta}\,
\3nab^{2}\Psi\;,\\
\dot{C}=&&K\left(\mathcal{A}_{C}\,C+ \mathcal{B}_{C}\,\Delta+
\mathcal{C}_{C}\,\Psi\right)+ \3nab^{2}\left(\mathcal{D}_{C}\, C
+\mathcal{E}_{\Delta}\Delta+ \mathcal{F}_{C}\,
\Psi\right)\;,\\
\nn\dot{\Psi}=&&\mathcal{A}_{\Psi}\,C+ \mathcal{B}_{\Psi}\,\Delta+
\mathcal{C}_{\Psi}\,\Psi+ \mathcal{D}_{\Psi}\, \3nab^{2}C
+\mathcal{E}_{\Psi}\,\3nab^{2}\Delta+ \mathcal{F}_{\Psi}\,
\3nab^{2}\Psi\;.
\end{eqnarray}
where for sake of simplicity the time dependent coefficients have
been indicated with the curly letters $\mathcal{A}$,...
$\mathcal{F}$. Expression for these coefficients are given
in the appendix. Note that as in the minimally coupled
case, the evolution of either the momentum perturbation $\Psi$ or the density
perturbation $\Delta$  is closed at second order (in time
derivatives).

\subsection{Harmonic decomposition}
(\ref{eq:SysPertFin}) is a system of partial differential
equations which is far too complicated to be solved directly.
For this reason we follow the standard procedure and perform
a harmonic decomposition of the perturbation equations
This operation is usually performed using eigenfunctions of the
Laplace-Beltrami operator on the 3-surfaces
of constant curvature that represent the homogeneous spatial
sections of the FLRW universes.

The situation is different in our approach: we only consider
quantities defined directly in the real perturbed universe and, in
general, the spatial derivative $\3nab_a$ is not a derivative on a
3-surface (unless the vorticity is zero  \cite{bi:EBH}). Therefore,
in order to use a harmonic decomposition we need to construct a
new set of harmonics using operators constructed from these derivatives
which are independent of proper time (i.e. constant on the fluid flow
lines). This is easily done by writing the Laplace-Beltrami operator
for $\3nab_a$ and defining our harmonics as the eigenfunction of
the Helmoltz equation (see \cite{bi:BDE} for more details):
\begin{eqnarray}\label{eq:harmonic}
  \3nab^{2}Q = -\frac{k^{2}}{a^{2}}Q \;,~~
  \3nab^{2}Q_a = -\frac{k^{2}}{a^{2}}Q_a \;,~~
  \3nab^{2}Q_{ab} = -\frac{k^{2}}{a^{2}}Q_{ab}\;,
\end{eqnarray}
where $k=2\pi a/\lambda$ is the wavenumber
and $\dot{Q}= \dot{Q}_a= \dot{Q}_{ab}=0$.

Using these harmonics we can expand every first order quantity in
the equations above,  for example \footnote{Note that the underlying
assumption in this decomposition is that the perturbation
variables can be factorised into purely temporal and purely
spatial components. \label{note:1}},
\begin{equation}\label{eq:developmentdelta}
\Delta=\sum \Delta_{(k)}(t)\;Q^{(k)}\mbox{}
\end{equation}
where $\sum$ stands for both a summation over a discrete index or
an integration over a continuous one. Using the harmonics $Q$, together
with equations (\ref{eq:harmonic}), the system (\ref{eq:SysPertFin})
can be written as
\begin{eqnarray}
\nn\dot{\Delta}_k=\left(\mathcal{A}_{\Delta} -
\frac{\mathcal{D}_{\Delta}\, k^2 }{a^2} \right)\,C_k +\left(
\mathcal{B}_{\Delta}-\frac{\mathcal{E}_{\Delta} \, k^2}{a^2}
\right)\,\Delta_k  +
  \left( \mathcal{C}_{\Delta}-\frac{\mathcal{F}_{\Delta} \, k^2}{a^2} \right) \,
   \Psi_k \label{eq:deltadot2SW}\;,
       \end{eqnarray}
       \begin{eqnarray}
\nn\dot{C}_k=\left( \mathcal{A}_{C} - \frac{\mathcal{D}_{C}\,
k^2}{a^2} \right)\, C_k +
  \left(\mathcal{B}_{C} - \frac{\mathcal{E}_{C}\, k^2}{a^2} \right)\, \Delta_k +
  \left(\mathcal{C}_{C} - \frac{\mathcal{F}_{C}\, k^2}{a^2} \right)\,
\Psi_k\label{eq:cdot2SW}\;,
       \end{eqnarray}
       \begin{eqnarray}
\nn\dot{\Psi}_k=\left( \mathcal{A}_{\Psi} -
\frac{\mathcal{D}_{\Psi}\, k^2}{a^2} \right)\, C_k +
  \left(\mathcal{B}_{\Psi} - \frac{\mathcal{E}_{\Psi}\, k^2}{a^2} \right)\, \Delta_k +
  \left(\mathcal{C}_{\Psi} - \frac{\mathcal{F}_{\Psi}\, k^2}{a^2} \right)\,
\Psi_k\label{eq:upsidot2SW}\;.
\end{eqnarray}
In this way we reduce the system to one of ordinary differential
equations that describe the evolution of the inhomogeneities for a
given scale and can be solved in the standard way, once the
background has been specified.
\subsubsection{The long wavelength limit $\lambda\gg H^{-1}$}

Before calculating the evolution of perturbations for specific FLRW
backgrounds it is worth considering an interesting sub-case. As we
can see from the equations (\ref{eq:harmonic}), by only considering
modes with wavelengths much larger than the the Hubble radius,
all the Laplacian terms can be neglected in (\ref{eq:SysPertFin}).
From the second equation in (\ref{eq:SysPertFin}), this implies
immediately that the curvature gradient $C_a$ and its divergence $C$
are conserved when the background universe is flat ($K=0$). In this
case the system (\ref{eq:SysPertFin}) can be simplified to give
\begin{eqnarray}\label{eq:sysLWL}
\dot{\Delta}=&\left.\mathcal{A}_{\Delta}\right|_{K=0}\,C+
\left.\mathcal{B}_{\Delta}\right|_{{K=0}}\,\Delta+
\left.\mathcal{C}_{\Delta}\right|_{K=0}\,\Psi\;,\\
\dot{\Psi}=&\left.\mathcal{A}_{\Psi}\right|_{K=0}\,C+
\left.\mathcal{B}_{\Psi}\right|_{K=0}\,\Delta+
\left.\mathcal{C}_{\Psi}\right|_{K=0}\,\Psi\;,
\end{eqnarray}
where $C$ is constant in time. Note that in the above
system, no spatial derivatives appear and as a consequence the
spatial part of $(\Delta,C,\Psi)$ can be factored out leaving a
second order system of ordinary differential equations. In the
treatment of the evolution of perturbation on super-horizon scales
and flat spatial geometry, we solve directly (\ref{eq:sysLWL}), using
for simplicity a plane wave expansion for the spatial part of the
perturbation variables. The long wavelength limit will be used
explicitly in the example given in the next section.

\section{Application: the case $F(\phi)=-\xi\phi^2$, $V(\phi)=\lambda\phi^{p}$}

Let us now consider a specific example using a background which may be
obtained, with a very natural choice of some integration
constants, in the context of Noether symmetries applied to the FLRW
metric \cite{bi:salvreview}. In this model the coupling function
is given by
\begin{equation}\label{couplingF}
 F(\phi) = - \xi (r) \phi^2,
\end{equation}
 and the potential
\begin{equation}
 V(\phi) = \lambda  \phi^{2 p(r)},
\end{equation}
where $\xi$ and $ p$  depend on a single free parameter $r$, via
\begin{equation}
 \xi (r) = \frac{(2r+3)^2}{12 (r+1)(r+2)} \quad ; \quad
 p(r) = \frac{3(1+r)}{3+2r}\;.
 \end{equation}
Equation (\ref{couplingF}) is different from the usual non-minimal
coupling $ F(\phi) = 1 + \zeta \phi^2 $ because in general it does
not contain the Hilbert Einstein term. In addition, the value of
$r$ determines the sign of the effective gravitational constant $
G_{eff}= F^{-1}$. Since the sign of the function $\xi(r)$ is not
always the same, there are, in general, values of $r$ for which
the gravitational interaction is effectively repulsive. These
facts deserve a detailed discussion that is outside the purpose of
this paper (in which this model is just a worked example) and will
be given in a forthcoming paper. However, roughly speaking, one
can imagine that the real coupling is actually $ F(\phi) = 1 - \xi
\phi^2 $ and that the condition (\ref{couplingF}) is realized only
at early times so that ordinary gravity is recovered at late
times.

Using the Noether symmetry approach with the above coupling and
potential leads to the solution
\begin{equation}\label{eq:NoethSol}
a(t) = \alpha t^n \quad ; \quad \phi(t) = \beta t^m ;\quad K=0\;,
\end{equation}
with
\begin{equation}
n(r) = \frac{2 r^2 + 9 r + 6}{r(r+3)} \quad ;
 \quad m(r) = - \frac{2r^2 + 9 r + 9}{r(r+3)}\;.
\end{equation}
which represents a specific case of a general exact analytic
solution for the background equations
\cite{Claudio,bi:salvreview,bi:marino}. Depending on the value of
$r$,  (\ref{eq:NoethSol}) can represent power law inflation
behaviour ($n(r)>1$) or a Friedmann-like phase
($0<n(r)<1$)\footnote{This solution  can represent, of course, also
a contraction, but in the following we will consider only the
expanding cases.} and its character is essentially related to the
choice of the form of the potential (via the choice of $r$). The
parameter $ \alpha $ is free while $ \beta $ is linked to $ \lambda
$ and $ r $ through
$$
\lambda = \frac{(6+r)(3+2r)^2}{8r(3+r)^2(2+3r+r^2)}
\beta^{-\frac{2r}{3+2r}}.
$$
It turns out simpler to leave $\beta$ free and derive $\lambda$,
but of course the real free parameter is this last one. Let us
write down the time dependence of some important quantities. The
expansion parameter is
$$
\Theta(t) = 3\frac{n(r)}{t};
$$
the effective energy density is
$$
\mu(t) = \frac{3(6+9r+2r^2)^2}{r^2(3+r)^2}\frac{1}{t^2},
$$
which is a non negative function of $r$ regardless of the sign of
$G_{eff}$; and the effective barotropic index is
$$
w=- \frac{2}{3}  -
  \frac{2 + r}
   {6 + r\,\left( 9 + 2\,r \right) },
$$
which is constant so that in this background the effective fluid
behaves like a barotropic fluid\footnote{This is a peculiar feature
of this background solution that is not true in in general. In
fact it can be easily verified that it is originated by the
peculiar relation between the time behaviour of the scalar field
and the power appearing in the potential of this example.}.

Let us now write the perturbation equations in the long wavelength
limit. As we have seen, in this limit the spatial dependence of
the perturbation variable can be considered as a plane wave and
can be factored out. Substituting the quantities above in the
perturbation equations (\ref{eq:sysLWL}) and in the constraint
(\ref{eq:DeltaUpsiC}) we obtain, after lengthy calculations,
\begin{eqnarray}\label{eq:sysNoetherFin}
 \dot{\Delta} &=&
   \mathcal{H}(r)\Delta+\mathcal{I}(r)\Psi+ \mathcal{M}(r)\frac{C}{{\alpha }^2}\,
  t^{-\frac{3\, \left( 1 + r \right) \, \left( 4 + r \right) }{r\,\left( 3 + r \right) }}\;,\label{eq:sysNoetherFin1}\\
\dot{\Psi} &=& \mathcal{N}(r)\Delta+\mathcal{O}(r)\Psi+
\mathcal{P}(r)\frac{C}{{\alpha }^2}\,
  t^{-\frac{3\, \left( 1 + r \right) \, \left( 4 + r \right) }{r\,\left( 3 + r \right)
  }}\;,\label{eq:sysNoetherFin2}\\ \dot{C}&=&0\\
\mathcal{Q}(r)\,\Delta&=&\mathcal{U}(r)\Psi+\mathcal{T}(r)\frac{C}{{\alpha
}^2}\, t^{-\frac{3\, \left( 1 + r \right) \, \left( 4 + r \right)
}{r\,\left( 3 + r \right) }}\;,\label{eq:sysNoetherFin3}
\end{eqnarray}
where
\begin{eqnarray}
\mathcal{H}(r) &=&  \frac{54 + r\,\left( 153 +  117\,r +
                    23\,r^{2} \right)   }{r\,{\left( 3 + r \right) }^2}\;, \\
 \mathcal{I}(r) &=& \frac{-2\,\left( 3 + 2\,r \right) \,\left( 6 + 5\,r \right) \,
     \left( 6 +  15\,r + 4\,r^{2} \right) }{r\,\left( 3 + r \right) \,
     \left( 6 + r\,\left( 9 + 2\,r \right)  \right) }\;,\\
 \mathcal{M}(r) &=&  -\frac{ r\,\left( 3 + r \right) \,\left( 6 + 5\,r \right) \,
       \left[21 + 2\,r\,\left( 21 + 12\,r + 2r^{2} \right)  \right]  }
       {6\,{\left( 6 + r\,\left( 9 + 2\,r \right)  \right) }^2}\;, \\
 \mathcal{N}(r) &=&  \frac{3\,\left( 6 + 9\, r + 2\,r^{2}  \right) \,
     \left( 6 + 15\,r + 4\,r^{2} \right) }{2\,r\,{\left( 3 + r \right) }^3}\;, \\
 \mathcal{O}(r) &=& -\frac{3\,\left[ 12 + r\,\left( 46 +  36\,r + 7\,r^{2} \right)
       \right] }{r\,{\left( 3 + r \right) }^2\,t}\;, \\
 \mathcal{P}(r) &=&  -\frac{ r\,\left[ 36 + r\,
          \left( 90 + 57\,r + 10\,r^{2}  \right)  \right]  }{12\,
     \left( 6 + r\,\left( 9 + 2\,r \right)  \right) }\;,\\
 \mathcal{Q}(r) &=&  3\,\left( 6 + r\,
     \left( 9 + 2\,r \right)  \right)\;,\\
 \mathcal{T}(r) &=&\frac{1}{6}r^2\,{\left( 3 + r \right) }^
     3\,\left( 3 + 2\,r \right)\;,  \\
 \mathcal{U}(r) &=&2\,\left( 3 + r \right) \,
  \left( 6 + 5\,r \right)\;.
\end{eqnarray}
Using the constraint (\ref{eq:sysNoetherFin3}), the above system can
be reduced to a particularly simple closed equation for $\Psi$:
\begin{equation}\label{eq:UpsiNoetherFin}
\dot{\Psi}=- \frac{\left( 6 + r \right)  }{\left( 3 + r \right)
\,t}\,\Psi-\frac{r\, \left( 3 + r \right) }{12}\,\frac{C}{{\alpha
}^2}\,t^
      {-\frac{3\, \left( 1 + r \right) \, \left( 4 + r \right) }
      {r\,\left( 3 + r \right) }}\;,
\end{equation}
where $C$ is constant. Note that in the above equation the role of
the spatial curvature perturbations C depends strictly on the value
of $r$. In particular, the $C$ term can grow or decrease in time
leading to a different effect on the overall perturbation dynamics.

Solving (\ref{eq:UpsiNoetherFin}) and using the constraint
(\ref{eq:DeltaUpsiC}), we obtain  the general solutions for $\Psi$
and $\Delta$
\begin{eqnarray}\label{eq:exactsols}
\Psi &=& \frac{r^2\,\left( 3 + r \right)^2}{12\,
          \left( 12 + 6\,r + r^2 \right) }\; \frac{C}{{\alpha }^2}\,
          t^{-2\,\frac{6 + 6\,r + r^2}{r\,\left( 3 + r \right) }}+
       \Psi_0 t^{-\frac{6 + r}{3 + r}}\;,\\
\nn\Delta &=& \frac{r^2\,\left( 3 + r \right)^{2}\,\left(12 + r (21
+ 2 r (6 + r))\right)}{18\, \left( 12 + r\,\left( 6 + r \right)
\right) \,
         \left( 6 + r\,\left( 9 + 2\,r \right)  \right)}\; \frac{C}{{\alpha }^2}\,
          t^{-2\,\frac{6 + 6\,r + r^2}{r\,\left( 3 + r \right) }}\\&+&
  \Psi_0 \,\frac{2\,\left( 3 + r \right) \,\left( 6 + 5\,r \right) }
    {3\,\left( 6 + r\,\left( 9 + 2\,r \right)  \right)}t^{-\frac{6 + r}{3 + r}}\;,
\end{eqnarray}
where $\Psi_0$ is an integration constant.

The detailed properties of the physics associated with these
solutions will be discussed in detail in a following paper. Here we
will limit ourselves to making  few brief comments. First of all in
the above solution  $\Delta$ has the same time dependence of $\Psi$
i.e. the scalar field clumps in the same way as the effective fluid.

\begin{figure}
  \includegraphics[width=13cm]{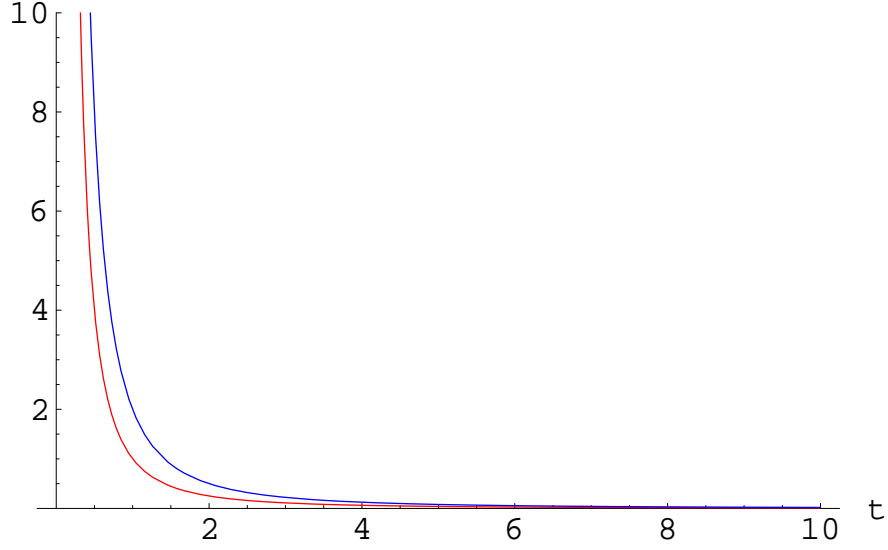}\\
  \caption{Plot of the solutions (\ref{eq:exactsols}) in the inflationary
  regime (case $r=0.01$). The solid curve represents the
  $\Psi/\Psi_0$ the dashed one represents $\Delta/\Psi_0$.
 }\label{fig:plotdeltaupsi1}
\end{figure}

\begin{figure}
  \includegraphics[width=13cm]{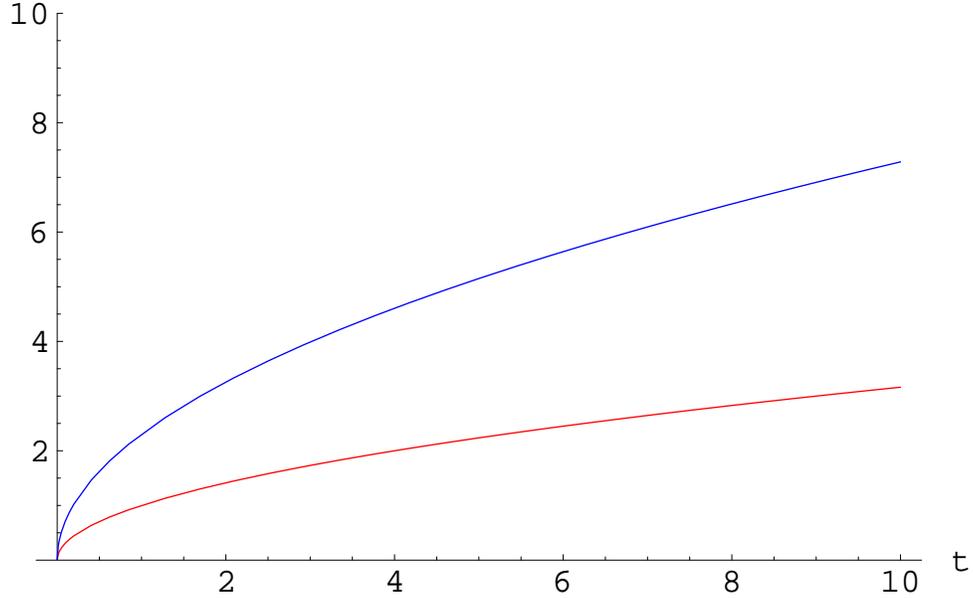}\\
  \caption{Plot of the solutions (\ref{eq:exactsols}) for the case $r=-5$. The solid curve represents the
  $\Psi/\Psi_0$ the dashed one represents $\Delta/\Psi_0$. }\label{fig:plotdeltaupsi5}
\end{figure}

In the inflationary regime ($ r < -3 - {\sqrt{3}}$, $-3 < r < -3 +
{\sqrt{3}}$, $r > 0$) the long wavelength perturbations are frozen
and decay (see Figure \ref{fig:plotdeltaupsi1}), unless
$-6<r<-3-\sqrt{3} $  for which the second mode of
(\ref{eq:exactsols}) is  growing (see Figure
\ref{fig:plotdeltaupsi5}).
The consequence of a growing mode on super-horizon scales during
inflation suggests that there is a deviation from scale invariance
for large $k$. As a consequence, any observed deviation from scale
invariance on these scales could be used to constrain these models.

If we consider a Friedmann regime $(-4.73 < r < -3.69 , -1.27 < r <
-0.81)$, the perturbations on large scale grow as if the effective
fluid was a form of standard matter. For example, supposing that the
expansion of the background follows the standard behaviour $a\propto
t^{2/3}$, which corresponds to $r\approx -4.17$, the parameter
$p(r)$ of the scalar field is $\approx 3.6$, and in this case
$w\approx 0$. If we choose $r=- 4$, we would have a radiation-like
effective fluid etc. This is means that a non-minimally coupled
scalar field in a Friedmann regime can mimic baryonic fluids, but
do not necessarily interact with photons via Thompson scattering.

Another curious behaviour in the Friedmann regime is that, for
$(-4.73< r <-4.17   , -1.07 < r < -1.27)$, the perturbations are
effectively frozen. This feature is very important when one makes
the transition from perturbations seeded in the early universe to
the matter power spectrum (if we take $\phi$ to correspond to the
inflaton), and in general will affect the spectrum of scalar
perturbations in the matter dominated era.

\begin{figure}
  \includegraphics[width=13cm]{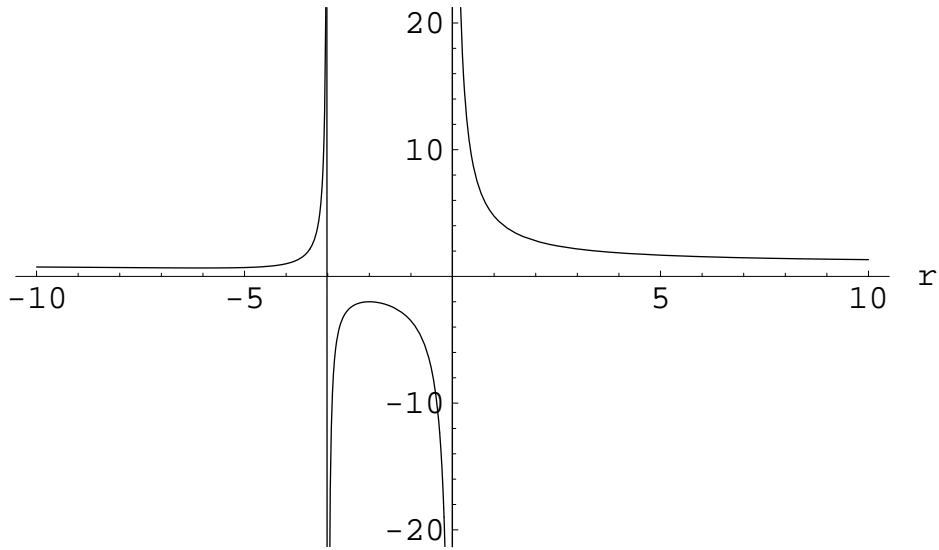}\\
  \caption{Plot of the exponent of the non-constant mode in (\ref{eq:Bardeen}) }\label{fig:bardeen}
\end{figure}

Finally, from the above equation we can also obtain the behaviour of
the Newtonian potential $\Phi_N$:
\begin{eqnarray}\label{eq:Bardeen}
    \Phi_N \nn &=& \frac{r^2\,\left( 3 + r \right)^{3}\,\left( 6 + 5\,r
            \right)}{18\, \left( 12 + r\,\left( 6 + r \right)  \right) \,
         \left( 6 + r\,\left( 9 + 2\,r \right)  \right)}\; C\\&+&
  \frac{2}{3}\;\frac{\left( 3 + r \right) \,\left( 6 + 5\,r \right) }
    { 6 + r\,\left( 9 + 2\,r \right)  }\;{\alpha }^2\;  \Delta_0\;t^{\frac{12+6r + r^{2}}{r(3 +
    r)}}\;.
\end{eqnarray}
As in the minimally coupled case, $\Phi$ has a constant mode, but
unlike the minimally coupled case, the other mode is not necessarily
decreasing. In fact, the exponent $\displaystyle{\frac{12+6r +
r^{2}}{r(3 + r)}}$ is negative only if $-3<r<0$ (see Figure
\ref{fig:bardeen}). This means that, when $r$ is outside the above
interval, the gravitational potential increases in time  so that the
rate at which the perturbations grow is faster than in the case for
a minimally coupled scalar field.

\section{Summary and Conclusions} \label{sec:ch7sum}

In this paper we have developed the theory of cosmological
perturbations of a  FLRW background for a generic scalar-tensor
theory of gravity in vacuum. Our analysis is based on the fact that
the field equations of a scalar-tensor theory  in vacuum can be
recast in a form that resembles the standard Einstein theory, in
which an effective fluid is present. This effective fluid is perfect
in the background (like in the case of a minimally coupled scalar
field), but it is imperfect in the perturbed universe. This is due
to the peculiar form of the heat flux and the anisotropic pressure
that depend on the acceleration and the shear as a result of the
non-minimal coupling $F$.

As in \cite{bi:peterscalar}, given our choice of $u^a$, the
perturbations can be characterized using the variables
$\dd_a=\mu^{-1} a \3nab_a\mu$, $C_a=a^{3} \3nab_a\tilde{R}$ and
$\Psi_a=\psi^{-1} a \3nab_a\psi$. The first one represents the
density perturbation of the effective fluid; the second one, the
projected gradient of the 3-curvature scalar of the $\phi=constant$
surfaces respectively; and the third one represents the
inhomogeneities of the scalar field. Note that, due to the fact that
in our frame $\3nab_a\phi=0$, this variable depends on the momentum
$\psi$.

The three variables above are connected to each other via the
constraint (\ref{eq:constr}). In the minimally coupled case such
constraint reduce to a proportionality relation between $\dd_a$  and
$\Psi_a$ and is used to eliminate the variable $\Psi_a$. However, in
our case $\Psi_a$ and $\dd_a$ have a rather different meaning and
different relevance depending on the problem one wishes to solve.
There can be cases in which the the perturbations of the effective
fluid have a different impact on the matter perturbation and it is
more useful to deal with $\dd_a$ than $\Psi_a$.

Once we  have chosen the perturbation variables it is relatively
easy (although rather long!) to derive their evolution equations.
These equation can be simplified if we consider spacetimes that do
not differ too much from the background. In fact, in this case one
can use the a linearization procedure based on the structure of the
background to completely drop the interaction terms.

In this paper, we limited ourselves to analyze spherically symmetric
clumping. This has been done by considering only the projected
divergence of the variables above. The equations for these variables
describe the evolution of scalar perturbations in a general
scalar-tensor theory of gravity in vacuum. They are valid also for
any spatial geometry and can be used to analyse the perturbations in
any scalar\hs field\hs dominated universe models (not necessarily
inflationary).

As an example we used a coupling, a potential and a background
solution obtained in the framework of the Noether Symmetry approach
\cite{bi:salvreview}. We were able to solve the perturbation
equations exactly for this background in the long wavelength limit.
Our solution reveals that there exist power law inflation regimes in
which the scalar field perturbations grow instead of being
dissipated. Such a behaviour will affect the matter power spectrum,
and potentially break scale invariance. However, this conclusion
depends strictly on the behaviour of the perturbations on
sub-horizon scales, which requires a complete analysis of the
perturbation equations for all scales.

Also, we found that in ``Friedmann regimes" the effective fluid
associated with a non-minimally coupled scalar field is able to
mimic a standard perfect fluid that does not necessarily interact
with photons. The perturbations of this fluid grow for most of the
values of $r$ associated to a Friedmann evolution, but there are
also cases in which the super-horizon perturbations are effectively
frozen, i.e., their growth rate is slower than the expansion rate,
thereby differing significantly from the standard gravitational
instability picture.

Finally, we derived the evolution of the gauge invariant version of
the Newtonian potential $\phi_N$. As in the case of a minimally
couple scalar field, $\phi_N$ contains a constant mode. However,
unlike the minimally coupled case, it also admits a second mode that
is not necessarily decaying and this again modifies the standard
picture.

In conclusion, our preliminary results reveal that the presence of a
non-minimal coupling has unexpected and interesting effects on the
evolution of the long wavelength scalar perturbations. An analysis
of the full perturbation spectrum will clarify further on these
properties and will be presented in a forthcoming paper.


\section{Acknowledgements} This work was supported by the National Research
Foundation (South Africa) and the {\it Ministrero deli Affari
Esteri- DIG per la Promozione e Cooperazione Culturale} (Italy)
under the joint Italy/South Africa science and technology
agreement.

\appendix
\section{Coefficients of the perturbation
equations} For sake of simplicity in the main body of the paper we
give here the expressions of the coefficient of the equations
(\ref{eq:SysPertFin}). All the quantities appearing in these
coefficients are meant to be the background ones.
\begin{eqnarray}
 \nn\mathcal{A}_{\Delta}(t)&=&  \frac{9\,K\,\dot{F}^2\,
      \left( F\,\left( 1 + w \right) \,\mu  + \Theta \,\dot{F} \right) }{2\,F\,a^4\,
      \left( 1 + w \right) \,\Theta \,{\mu }^2\,\left( 2\,F\,{\psi }^2 +3\,\dot{F}^2  \right) }
\\\nn &+&\frac{ 3\,F\,\left( 1 + w \right) \,\mu
            \,\psi^{2}  + \Theta \,\dot{F}\,\left( 2\,{\psi }^2 -3\,F\,\left( 1 + w
            \right) \,\mu   -2\,\Theta \,\dot{F}  + 3\,\ddot{F}\right)
            }{2\,a^2\,\Theta \,\mu \,\left( 2\,F\,{\psi }^2 +3\,\dot{F}^2  \right) }
\end{eqnarray}
\begin{eqnarray}
 \nn\mathcal{B}_{\Delta}(t)&=& \frac{K\,\dot{F}^2\,\left[
            3\,F\,\mu \,  \left( 2\,w\,{\Theta }^2 - 3\,\left( 1 + w \right) \,\mu  \right)  -
            6\,{\Theta }^2\,{\psi }^2 +
           \Theta \,\dot{F}\,\left( 2\,{\Theta }^2\, - 9\,\mu \right)  - 6\,{\Theta }^2\,\ddot{F} \right] }{F\,
            a^2\,\left( 1 + w \right) \,\Theta \,{\mu }^2\,\left( 2\,F\,{\psi }^2 +3\,\dot{F}^2  \right)}
\\ \nn &+&\frac{3\,F \,\psi^{2}\,
            \left( 2\,w\,{\Theta }^2 - 3\,\left( 1 + w \right) \,\mu  \right)
            + 2\,\Theta \,F\,\dot{F}\,\left( 2\,{\Theta }^2 + 6\,\mu  +
            9\,w\,\mu  \right) }{3\,\Theta  \,
             \left( 2\,F\,{\psi }^2 +3\,\dot{F}^2  \right) }
\\ &-&\frac{\Theta \,\dot{F}\,\left(
               6\,{\psi }^2 - 3\,\dot{F}\, \left( 1 + 3\,w \right) \,\Theta \,\psi
               +9\,\ddot{F} \right)  }{3\,\Theta  \,
               \left( 2\,F\,{\psi }^2 +3\,\dot{F}^2  \right) }
 \end{eqnarray}
\begin{eqnarray}
 \nn\mathcal{C}_{\Delta}(t)&=&\frac{6\,\Theta \,
            \left( {\Theta }^2 + \mu  + 3\,w\,\mu  \right) \,\dot{F}^{2}
            - \left( 4\,{\Theta }^2 + 3\,\mu  + 9\,w\,\mu \right)
            \,\dot{F}\psi^{2}
             - 6\,\Theta \,\ddot{F}\psi^{2} }{3\,\mu\left( 2\,F\,{\psi }^2 +3\,\dot{F}^2  \right)}
\\ &-& \frac{2\,K\,\Theta  \,\dot{F}^2\,
      \left[ 4\,\Theta \,\dot{F} - 3\,\psi^{2} \,\left( 2 + F'' \right)  \right] }{F\,a^2\,
      \left( 1 + w \right) \,{\mu }^2\,\left( 2\,F\,{\psi }^2 +3\,\dot{F}^2  \right) }
 \end{eqnarray}
\begin{eqnarray}
  \mathcal{D}_{\Delta}(t)&=&-\frac{\dot{F}^3 }
   {2\,{\mu }^2\,F\,a^2\,\left( 1 + w \right) \,\left( 2\,F\,{\psi }^2 +3\,\dot{F}^2  \right) }\nn
  \end{eqnarray}
\begin{eqnarray}
  \mathcal{E}_{\Delta}(t)&=&\frac{ \dot{F}^{3}\,\left( 2\,{\Theta }^2  + 3\,\mu\right)
     + 6\,\Theta\,\dot{F}^{2} \,\left( F\,w\,\mu  - {\psi }^2 - \ddot{F}\right)
     }{3\,F\,\left( 1 + w \right) \,{\mu }^2\,\left( 2\,F\,{\psi }^2 +3\,\dot{F}^2  \right) }\nn
  \end{eqnarray}
\begin{eqnarray}
 \nn \mathcal{F}_{\Delta}(t)&=&\frac{6\,F^2\,\left( 1 + w
\right) \,\mu \,\dot{F}\,\psi^{2}  +
       6\,F\,\Theta\,\dot{F}^{2}\,\left( 2\,{\psi }^2  - F\,\left( 1 + w \right) \,\mu\right)}{3\,F^2\,
     \left( 1 + w \right) \,{\mu }^2\,\left( 2\,F\,{\psi }^2 +3\,\dot{F}^2  \right) }
\\ &+& \frac{\dot{F}^{3}\,\left(9\,F\,\left( 1 + w \right) \,\mu  + 12\,\Theta \,\dot{F}-8\,F\,{\Theta }^2  \right)
+6\,F\,\Theta\dot{F}^{2} \,\psi^{2} \,F'' }{3\,F^2\,
     \left( 1 + w \right) \,{\mu }^2\,\left( 2\,F\,{\psi }^2 +3\,\dot{F}^2  \right) }
 \end{eqnarray}
\begin{eqnarray}
 \mathcal{A}_C(t)&=& \frac{ 3F\,\mu\,\left( 1 + w \right) +
3\Theta \,\dot{F}}{\mu\,F\,a^2\,\Theta \,\left( 1 + w \right)  }
 \end{eqnarray}
\begin{eqnarray}
 \mathcal{B}_C(t)&=& \frac{ 6\,F\,\mu \,\left( 2\,w\,{\Theta
}^2 - 3\,\left( 1 + w \right) \,\mu  \right)  - 12\,{\Theta
}^2\,{\psi }^2 + 2\Theta \,\dot{F}\, \left( 2\,{\Theta }^2  -
9\,\mu  \right)  -12\,{\Theta }^2\,\ddot{F}}{3\,F\,\left( 1 + w
\right) \,\Theta \,\mu }
\end{eqnarray}
\begin{eqnarray}
 \mathcal{C}_C(t)&=& \frac{4\,\Theta \,
     \left( 3\,\psi^{2} \,\left( 2 + F'' \right)  -4\,\Theta \,\dot{F}\right) }{3 \,\mu\,F\,
     \left( 1 + w \right) }
\end{eqnarray}
\begin{eqnarray}
  \mathcal{D}_C(t)&=& -\frac{\psi \,F'}{3\,F\,\mu \left( 1 + w \right)}
\end{eqnarray}
\begin{eqnarray}
 \mathcal{E}_C(t)&=&\frac{ 2\,a^2\,\dot{F}\,\left( 2\,{\Theta
}^2  + 3\,\mu \right) + 12\,a^2\,\Theta \,\left( F\,w\,\mu  -
{\psi }^2 - \ddot{F} \right) }{9\,\mu \,F\,\left( 1 + w \right) }
\end{eqnarray}
\begin{eqnarray}
 \mathcal{F}_C(t)&=&\frac{2\,a^2 \, \left[ F\,\dot{F}\,\left(
9\,\left( 1 + w \right) \,\mu  -8\,{\Theta }^2\right)  +
12\,\Theta \,\dot{F}^2 + 6\,F\,\Theta \,\psi^{2} \,\left( 2 + F''
\right)  \right] }{9 \,\mu\,F^2\,\left( 1 + w \right) }
\end{eqnarray}
\begin{eqnarray}
 \mathcal{A}_{\Psi}(t)&=&\nn -\frac{9\,K\,\dot{F}\,\left(
F\,\left( 1 + w \right) \,\mu
        + \Theta \,\dot{F} \right) }{2\,a^4\,\left( 1 + w \right) \,
      {\Theta }^2\,\mu \,\left( 2\,F\,\psi^{2} + 3\,\dot{F}^2 \right) }
 \\ &+&  \frac{ 6\,\mu\,\Theta\,F^2\,\left( 1 + w \right) +  \dot{F}\,
            \left[ F\left(4\,{\Theta }^2 + 9\,\left( 1 + w \right) \,\mu\right)  +
              6\,\Theta \,\dot{F}\right]  -
        6\,F\,\Theta \,\ddot{F}  }{4\,a^2\,{\Theta }^2\,
      \left( 2\,F{\psi }^2 + 3\,\dot{F}^2 \right) }
\end{eqnarray}
\begin{eqnarray}
 \mathcal{B}_{\Psi}(t)&=&\nn -\frac{ 4\,\mu\,F^2\,\Theta
\,\left( 2\,{\Theta }^2 + 6\,\mu +
                9\,w\,\mu  \right)- 18\,\mu\,F\,\Theta \,\ddot{F}
                }{6\,{\Theta }^2\,\left( 2\,F{\psi }^2 + 3\,\dot{F}^2 \right) }
\\&+&\nn \frac{3\,\mu\,\dot{F} \,\left( 2\,F\,{\Theta }^2 +
                9\,F\,\left( 1 + w \right) \,\mu  + 6\,\Theta \,\psi \,F' \right)
        }{6\,{\Theta }^2\,\left( 2\,F{\psi }^2 + 3\,\dot{F}^2 \right) }
  \\\nn&-&K\left\{\frac{\,F'\,\left( 3\,F\,\mu \,
         \left( 2\,w\,{\Theta }^2 - 3\,\left( 1 + w \right) \,\mu  \right)  -
        6\,{\Theta }^2\,{\psi }^2 \right) }{a^2\,
      \left( 1 + w \right) \,{\Theta }^2\,\mu \,\psi \,\left( 2\,F + 3\,{F'}^2 \right)
      }\right.
\\ &+&\left.\frac{\Theta \,\dot{F}^{2}\,\left( 2\,{\Theta }^2\,\psi  - 9\,\mu \,\psi \right)
        - 6\,F'\,{\Theta }^2\,\ddot{F}
       }{a^2\, \left( 1 + w \right) \,{\Theta }^2\,\mu  \,\left( 2\,F\,\psi^{2} + 3\,\dot{F}^2
       \right)}\right\}
    \end{eqnarray}
\begin{eqnarray}
 \mathcal{C}_{\Psi}(t)&=&\nn\frac{2\,K\,\dot{F}\,
      \left( 4\,\Theta \,\dot{F} - 3\,\psi^{2} \,\left( 2 + F'' \right)  \right) }{a^2\,
      \left( 1 + w \right) \,\mu \,\left( 2\,F\,\psi^{2}  + 3\,\dot{F}^2 \right) }
  \\ &-&\nn\frac{F\, \psi^{2}\,\left[ 2\,{\Theta }^2 - 3\,\left( \mu  + 3\,w\,\mu
            \right) \right]   + 3\,\Theta \,\left( 4\,F\,\psi\,\psi ' +
           3\,\dot{F}^2\,\Theta \right) }{3\,\Theta  \,
      \left( 2\,F\, \psi^{2} + 3\,\dot{F}^2 \right) }
  \\  &+&
           \frac{3\,\Theta \,\dot{F}\,\left( 2\,F\,\left( {\Theta }^2 + \mu  + 3\,w\,\mu  \right)  +
              3\,\ddot{F} \right) }{3\,\Theta \,
      \left( 2\,F\,\psi^{2} + 3\,\dot{F}^2 \right) }
\end{eqnarray}
\begin{eqnarray}
  \mathcal{D}_{\Psi}(t)&=&\frac{\dot{F}^2}
   {2\,\mu\,a^2 \,\Theta  \,\left( 1 + w \right) \,\left( 2\,F\,{\psi }^2 +3\,\dot{F}^2  \right)}
  \end{eqnarray}
\begin{eqnarray}
   \mathcal{E}_{\Psi}(t)&=& \frac{6\,\Theta \,\dot{F}\,\left( {\psi
        }^2 - F\,w\,\mu +\ddot{F} \right)-\dot{F}^{2}\,\left( 2\,{\Theta }^2  + 3\,\mu \right)}{
        3 \,\Theta
        \,\mu \,\left( 1 + w \right)\,\left( 2\,F\,{\psi }^2 +3\,\dot{F}^2  \right) }
\end{eqnarray}
\begin{eqnarray}
\mathcal{F}_{\Psi}(t)&=&  \frac{2\,\dot{F}\,
     \left[ 3\,F\,\left( F\,\left( 1 + w \right) \,\mu  - 2\,{\psi }^2 \right)  +
        4\,F\,\Theta \,\dot{F} - 6\,\dot{F}^2 - 3\,F\,\psi^{2} \,F''
       \right] }{3\,\mu\,F\,\left( 1 + w \right)  \,\left( 2\,F\,{\psi }^2 +3\,\dot{F}^2  \right) }
\end{eqnarray}

\end{document}